\documentclass[journal=jctc,manuscript=article]{achemso}

\usepackage[version=3]{mhchem} 
\usepackage{graphicx}
\usepackage{amsmath}
\usepackage{amssymb}
\usepackage{color}
\usepackage[normalem]{ulem}
\usepackage{chemformula}
\usepackage{float}
\usepackage[symbol]{footmisc}

\SectionNumbersOn

\author{Wei Fang}
\alsoaffiliation{Laboratory of Physical Chemistry, ETH Z\"urich, 8093 Z\"urich, Switzerland}
\affiliation{Department of Chemistry, Fudan University, Shanghai 200438, P. R. China}
\alsoaffiliation{State Key Laboratory of Molecular Reaction Dynamics and Center for Theoretical Computational Chemistry, Dalian Institute of Chemical Physics, Chinese Academy of Sciences, Dalian 116023, P. R. China.}
\email{wei_fang@fudan.edu.cn}

\author{Yu-Cheng Zhu}
\affiliation{State Key Laboratory for Artificial Microstructure and Mesoscopic Physics, Frontier Science Center for Nano-optoelectronics and School of Physics, Peking University, Beijing 100871, China}

\author{Yi-Han Cheng}
\affiliation{State Key Laboratory for Artificial Microstructure and Mesoscopic Physics, Frontier Science Center for Nano-optoelectronics and School of Physics, Peking University, Beijing 100871, China}

\author{Yi-Ping Hao}
\affiliation{State Key Laboratory of Molecular Reaction Dynamics and Center for Theoretical Computational Chemistry, Dalian Institute of Chemical Physics, Chinese Academy of Sciences, Dalian 116023, P. R. China.}

\author{Jeremy O. Richardson}
\affiliation{Laboratory of Physical Chemistry, ETH Z\"urich, 8093 Z\"urich, Switzerland}

\title{Robust Gaussian Process Regression method for efficient reaction pathway optimization: application to surface processes}

\begin{document}



\newpage
\begin{abstract}
Simulation of surface processes is a key part of computational chemistry that offers atomic-scale insights into mechanisms of heterogeneous catalysis, diffusion dynamics, as well as quantum tunneling phenomena.
The most common theoretical approaches involve optimization of reaction pathways, including semiclassical tunneling pathways (called instantons).  However, the computational effort can be demanding, especially for instanton optimizations with \textit{ab initio} electronic structure.
Recently, machine learning has been applied to accelerate reaction-pathway optimization, showing great potential for a wide range of applications.
However, previous methods suffer from practical issues such as unfavorable scaling with respect to the size of the descriptor, and were mostly designed for reactions in the gas phase.
We propose an improved framework based on Gaussian process regression for general transformed coordinates, which can alleviate the size problem. The descriptor combines internal and Cartesian coordinates, which improves the performance for modeling surface processes.
We demonstrate with eleven instanton optimizations in three example systems that the new approach makes \textit{ab initio} instanton optimization significantly cheaper, such that it becomes not much more expensive than a classical transition-state theory calculation. 
\end{abstract}

\newpage
\section{Introduction}
Surface processes and reactions on surfaces are at the core of many important phenomena, such as heterogeneous catalysis, ice nucleation, and corrosion, to name just a few.
For a long time, surface science has been the focus of an enormous amount of experimental and theoretical research.
Computer simulations have been essential for understanding surface structure, reactions and dynamics on surfaces, as well as quantum tunneling phenomena in these systems, achieving tremendous success \cite{surface_simulation_review,Water_at_Interfaces,Stamatakis2012}. 
Locating and optimizing reaction pathways is one of the most crucial parts of modern computational chemistry, offering insights for the mechanism of surface processes and reactions \cite{CI_NEB}.
Reaction rates can also be estimated from the transition states (TSs) based on classical transition-state theory (TST) \cite{Eyring1935review}.
As the demand of higher accuracy modeling increases, incorporation of nuclear quantum effects (NQEs), in particular quantum tunneling, in simulations is becoming the new standard.
Ring-polymer instanton theory is a robust method for rigorously including tunneling effects into the simulation of reaction rates and mechanisms \cite{Miller1975rate,Andersson-2009-JPCA-4468,RPInst,Kastner-2014-WIRE,Inst_persp,InstReview}, with a good balance between accuracy and efficiency.
It employs a semiclassical approximation to define a dominant tunneling pathway called an instanton, which can be located via a first-order saddle point optimization on the ring-polymer potential energy surface.
It is important to note that this pathway is not in general equivalent to the minimum-energy pathway.
The instanton rate only requires local properties (i.e. potentials, gradients, and Hessians) along the tunneling pathway (instanton), due to the semiclassical approximation.
Thus the ring-polymer instanton theory can be viewed as a quantum-mechanical extension of classical TST.
Using instanton theory, recent studies have unveiled several interesting phenomena on different surfaces related to quantum tunneling \cite{Jonsson_2011,Erlend_Activation_2014,broadtop,Fang2020,Fang2022,Litman2020,Cheng_2022},
demonstrating the importance of modeling quantum tunneling in the simulation of surface processes.
Although instanton theory is much more efficient than a full quantum calculation, it is still more demanding than classical TST,
especially when combined with \textit{ab initio} electronic structure calculations, which has so far impeeded the wide application of rigorous tunneling calculations.
Therefore, reducing the cost of the optimization of reaction pathways such as instantons is important to the future development of this field.
%

Earlier works on improving optimization schemes were mostly dedicated to 
finding better coordinate systems in which to perform the optimization\cite{delocalizedInternals,Fogarasi_1992,Baker_1996,Billeter2000,Paizs2000,Nemeth2010},
or on improving the approximate Hessian of the system 
to accelerate convergence \cite{John_1986,Bofill1994update}.
The last decade has witnessed a blossom of the application of machine-learning methods in computational chemistry, which started a revolution in the field. 
In recent years, machine learning methods have been applied to various optimization problems \cite{Peterson_saddle_2016,Jonsson_GPR_2016,Jonsson_GPR_2017,Laude2018,Kastner_PES_NN_rate_2018,Hauser_GPR_2020,Kastner_GPR_2021}, challenging the conventional algorithms that have stood for decades.
These methods use machine learning to fit the local potential-energy surface (PES) around a local minimum geometry or the dominant reaction/tunneling pathway and perform optimizations on the fitted PES. 
By iterating this procedure they can be converged to give the same pathway as for the true PES at a fraction of the cost.
For instanton optimization, in addition to the learning and prediction of energies and gradients that are commonly done with machine learning, Hessian training is also crucial \cite{Laude2018}.
Methods based on both neural networks (NN) and Gaussian process regression (GPR) have been explored.
The predictive performance of GPR has been shown to be very competitive \cite{Csanyi_tungsten_2014}, especially when the training data size is small \cite{nnreview2001}, therefore it is preferred for this application.
Note that this application is significantly different from fitting a global PES with machine learning \cite{Schmidt2019,Csanyi_tungsten_2014}, 
because we require high accuracy only in a small local region.
For this reason, we can limit ourselves to a much smaller set of training data.
%

Despite the great success of GPR-assisted optimization methods shown in the previous works, there are still several issues that obstruct the wide application of this method, especially for application to surface systems.
A major issue with the previous GPR optimization methods \cite{Laude2018} is that since they were mostly designed for reactions in the gas phase (which requires translational and rotational invariance), the descriptors used, namely internal coordinates, do not work well for surface systems.
In addition there are a number of practical issues.
For example, the use of delocalized internals in GPR for planar molecules may cause numerical problems.
Also, memory and efficiency issues caused by e.g. Hessian training or the use of long descriptors (such as redundant internals) can plague the performance in larger systems.
Therefore, more methodological developments are urgently needed for the maturation of GPR optimization methods.
%

In this work, we develop an improved GPR method for geometry optimization that: (i) is suitable for modeling surface reactions and processes; (ii) can be applied to instanton optimization (i.e. it includes hessian training); (iii) addresses previous practical issues.
We test the performance of our method on three test systems each covering a different type of surface:
(a) \ce{H2O} dissociation on Cu(111), a representative surface catalysis reaction; (b) \ce{CH2O} rotation on Ag(110), a dynamical process on surface that can be observed in STM experiments; (c) double proton transfer (DPT) in the formic acid dimer (FAD) on NaCl(001), a reaction between adsorbates on surface.
Good performance is observed for GPR assisted instanton optimization in these test systems, showing fast convergence of \textit{ab initio} gradients with just few iterations.
Our also GPR alleviates some of the previous practical issues, making GPR optimizations more robust.
In particular, we show that ``selective Hessian training'' (available within the method introduced in this work) provides a computational advantage for the application to larger systems.
In particular, it converges significantly faster and requires far less computational effort than an instanton optimization carried out with conventional methods.
Nonetheless, the new approach retains the accuracy of the original instanton method as the results are formally identical once the algorithm is converged.

\section{Theory}
Given a training set of $M$ geometries \{\textbf{x}$_m$\} and potential energies \{$V(\textbf{x}_m)$\} (also called $V_m$ for short), the GPR prediction of the potential energy for a new geometry \textbf{x}$^*$ is
\begin{equation}
V(\textbf{x}^*)-\bar{V}=\sum_{m=1}^{M}k(\textbf{x}^*,\textbf{x}_m;\boldsymbol{\theta})w_m
\label{gpr0}
\end{equation}
where $\bar{V}$ is the average potential of the training set, $k$ is the kernel function, $\boldsymbol{\theta}$ is a vector of hyperparameters, and $\textbf{w}=(w_1~...~w_M)^T$ are the weights.
There are many possible choices for the kernel function \cite{WilliamsBook}, but in this work, we use the Gaussian kernel $k(\textbf{x}_i,\textbf{x}_j)=\theta_1\exp (-\theta_2||\textbf{x}_i-\textbf{x}_j||^2)$.
The weights are determined by solving a set of linear equations $(\textbf{K}+\sigma^2\textbf{I}_M)\textbf{w}=(V_1~...~V_M)^T$, where \textbf{K} is a matrix with element $k(\textbf{x}_i,\textbf{x}_j)$ in the $i$-th row and $j$-th column, \textbf{I}$_M$ is the identity matrix of rank $M$, and $\sigma$ is the noise hyperparameter.

\subsection{GPR for gradient and Hessian learning}
For geometry optimization, learning and predicting gradients are important, while for instanton optimizations, Hessian learning and prediction becomes necessary in practice.
This can be achieved with an extension \cite{Jonsson_GPR_2017} to Eq. \ref{gpr0},
\begin{equation}
\begin{pmatrix}
V(\textbf{x}^*)-\bar{V}\\\left.\frac{\text{d}V}{\text{d}\textbf{x}}\right|_{\textbf{x}^*}\\\left.\frac{\text{d}^2V}{\text{d}\textbf{x}^2}\right|_{\textbf{x}^*}
\end{pmatrix}
=
\begin{pmatrix}
\textbf{k}_\text{ext,x}^T(\textbf{x}^*)\\
\frac{\text{d}\textbf{k}_\text{ext,x}^T(\textbf{x}^*)}{\text{d}\textbf{x}^*}\\
\frac{\text{d}^2\textbf{k}_\text{ext,x}^T(\textbf{x}^*)}{\text{d}{\textbf{x}^*}^2}
\end{pmatrix}
\textbf{w}_\text{ext}
\label{gpr_ex}
\end{equation}
For clarity and convenience, we will define a set of conventions here:
\begin{itemize}
    \item \textbf{bold lower case}: column vector.
    \item \textbf{bold upper case}: matrix.
    \item $\frac{\text{d}V}{\text{d}\textbf{x}}$: column vector of length $f_\text{x}$ ($f$ denotes the length of the vector noted in the subscript).
    \item $\frac{\text{d}^2V}{\text{d}\textbf{x}^2}$: column vector of length $f_\text{Hx}=f_\text{x}(f_\text{x}+1)/2$. This is the upper triangle of the Hessian matrix, flattened into an array in the row-major ordering.
    \item $\frac{\text{d}\textbf{k}}{\text{d}\textbf{x}}$: Matrix with $f_\text{x}$ rows and $f_\text{k}$ columns.
    \item $\frac{\text{d}^2\textbf{k}}{\text{d}\textbf{x}^2}$: Matrix with $f_\text{Hx}$ rows and $f_\text{k}$ columns.
\end{itemize}
Also, we order the training data such that the first $M_g$ entries have gradient data, and the first $M_H<=M_g$ entries additionally have Hessian data.
Within this notation the terms in Eq.~\ref{gpr_ex} can be written as,
\begin{equation}
\textbf{k}_\text{ext,x}(\textbf{x}^*)=
\begin{pmatrix}
k(\textbf{x}^*,\textbf{x}_1);~...~;~k(\textbf{x}^*,\textbf{x}_M);~
\frac{\text{d}k(\textbf{x}^*,\textbf{x}_1)}{\text{d}\textbf{x}_1};~...~;~\frac{\text{d}k(\textbf{x}^*,\textbf{x}_{M_g})}{\text{d}\textbf{x}_{M_g}};~ 
\frac{\text{d}^2k(\textbf{x}^*,\textbf{x}_1)}{\text{d}\textbf{x}_1^2};~...~;~\frac{\text{d}^2k(\textbf{x}^*,\textbf{x}_{M_H})}{\text{d}\textbf{x}_{M_H}^2}
\end{pmatrix}
\label{kextx}
\end{equation}
(the subscript $x$ denotes the kernel derivatives are taken with respect to Cartesian coordinates).
$\textbf{w}_\text{ext}$ is the extended weights, obtained by solving the following set of linear equations,
\begin{equation}
(\textbf{K}_{xx}+\boldsymbol{\Lambda}_{xx})\textbf{w}_\text{ext}=\textbf{y}_x
\label{model}
\end{equation}
in which
\begin{equation}
\textbf{K}_{xx}
\equiv
\begin{pmatrix}
\textbf{K} & \left(\frac{\text{d}\textbf{K}}{\text{d}\textbf{x}}\right)^T & \left(\frac{\text{d}^2\textbf{K}}{\text{d}\textbf{x}^2}\right)^T \\
\frac{\text{d}\textbf{K}}{\text{d}\textbf{x}'} & \frac{\text{d}}{\text{d}\textbf{x}'}\left(\frac{\text{d}\textbf{K}}{\text{d}\textbf{x}}\right)^T & \frac{\text{d}}{\text{d}\textbf{x}'}\left(\frac{\text{d}^2\textbf{K}}{\text{d}\textbf{x}^2}\right)^T \\
\frac{\text{d}^2\textbf{K}}{\text{d}\textbf{x}'^2} & \frac{\text{d}^2}{\text{d}\textbf{x}'^2}\left(\frac{\text{d}\textbf{K}}{\text{d}\textbf{x}}\right)^T & \frac{\text{d}^2}{\text{d}\textbf{x}'^2}\left(\frac{\text{d}^2\textbf{K}}{\text{d}\textbf{x}^2}\right)^T
\end{pmatrix}
=
\begin{pmatrix}
\textbf{k}_\text{ext,x}^T(\textbf{x}_1)\\
\vdots\\
\textbf{k}_\text{ext,x}^T(\textbf{x}_M)\\
\frac{\text{d}\textbf{k}_\text{ext,x}^T(\textbf{x}_1)}{\text{d}\textbf{x}_1}\\
\vdots\\
\frac{\text{d}\textbf{k}_\text{ext,x}^T(\textbf{x}_{M_g})}{\text{d}\textbf{x}_{M_g}}\\
\frac{\text{d}^2\textbf{k}_\text{ext,x}^T(\textbf{x}_1)}{\text{d}\textbf{x}_1^2}\\
\vdots\\
\frac{\text{d}^2\textbf{k}_\text{ext,x}^T(\textbf{x}_{M_H})}{\text{d}\textbf{x}_{M_H}^2}
\end{pmatrix}
\label{Kxx}
\end{equation}
is the extended covarience matrix, 
\begin{equation}
\boldsymbol{\Lambda}_{xx}=
\begin{pmatrix}
\sigma^2\textbf{I}_M &  &   \\
  & \sigma_g^2\textbf{I}_{M_gf_\text{x}} &  \\
 &  & \sigma_H^2\textbf{I}_{M_Hf_\text{Hx}}
\end{pmatrix}
\end{equation}
is the noise matrix, and
\begin{equation}
\textbf{y}_x=\left(V_1;~...~;V_M;~\frac{\text{d}V_1}{\text{d}\textbf{x}};~...~;\frac{\text{d}V_{M_g}}{\text{d}\textbf{x}};~\frac{\text{d}^2V_1}{\text{d}\textbf{x}^2};~...~;\frac{\text{d}^2V_{M_H}}{\text{d}\textbf{x}^2} \right)
\end{equation}
is the training data.

\subsection{GPR for general descriptors}
Constructing GPR potentials using the Cartesian coordinate is the simplest and most straight-forward way.
However, it has some drawbacks, for example, when describing gas phase molecules, the Cartesian coordinates are not translationally nor rotationally invariant.
Also, it is not a natural way of describing covalent bonds.
Since the descriptors are arguably the most important part of building a good machine learning potential \cite{Csanyi_PRB_2013,Schmidt2019}, it is desirable to design a GPR method for transformed descriptors \textbf{q}.
Such idea has been probed for certain coordinate transformations, such as the redundant and delocalized internal coordinates \cite{delocalizedInternals}.
The GPR method designed in previous works \cite{Laude2018,Kastner_GPR_2021} functions in the following manner: all observables (gradients and Hessians) are transformed from Cartesian coordinates into internal coordinates in the training process.
In the prediction process, it first predicts gradients and Hessians in internal coordinates, then one transform them back into Cartesian gradients and Hessians.
This approach has some drawbacks, specifically, the back and forth transformation from \textbf{x} to \textbf{q} may encounter numerical issues, for example, it breaks down for planar and linear molecules if the delocalized internals is used. 
A second approach has been explored in another work, which avoids the transformation of physical observables \cite{Hauser_GPR_2020}.
However, the alternative approach has only been developed for gradient training, which is applicable to geometry optimization, but is inadequate for instanton optimizations.
Here we adapt and extend the second approach to Hessian training and prediction. 
%

We define a general coordinate (descriptor) \textbf{q} as
\begin{equation}
\textbf{q} = \textbf{J}~\textbf{x}
\end{equation}
where $\textbf{J}$ is the Jacobian matrix
\begin{equation}
\textbf{J}\equiv\textbf{J}_{qx}(\textbf{x})=
\begin{pmatrix}
\frac{\text{d}q_1}{\text{d}x_1} & \dots & \frac{\text{d}q_1}{\text{d}x_{f_x}} \\
\vdots & & \vdots \\
\frac{\text{d}q_{f_q}}{\text{d}x_1} & \dots & \frac{\text{d}q_{f_q}}{\text{d}x_{f_x}} \\
\end{pmatrix}
\end{equation}
The kernel function in the \textbf{q} coordinate representation is 
\begin{equation}
k(\textbf{q}_1,\textbf{q}_2;\boldsymbol{\theta})=\theta_1\exp (-\theta_2||\textbf{q}_1-\textbf{q}_2||^2).
\label{kernel}
\end{equation}
When the kernel \textbf{K} is defined using the transformed coordinates \textbf{q}, the main difficulty is the construction of the extended covariance matrix \textbf{K}$_{xx}$.
We define a transformation matrix \textbf{L}, where $\textbf{K}_{xx} = \textbf{L}\textbf{K}_{qq}\textbf{L}^T$.
$\textbf{K}_{qq}$ is defined similarly as $\textbf{K}_{xx}$, but with the kernel derivatives taken with respect to \textbf{q} instead of \textbf{x}, which follows directly from differentiating Eq. \ref{kernel}.
\textbf{L} can be derived using the chain rule, and it can be formally written as,
\begin{equation}
\textbf{L}=
\begin{pmatrix}
\textbf{I}_M &  &  &  &  &  &  & \\
  & \textbf{J}^T(\textbf{x}_1) &  &  &  &  &  & \\
  &  & \ddots &  &  &  &  & \\
  &  &  & \ddots &  &  &  & \\
  &  &  &  & \textbf{J}^T(\textbf{x}_{M_g}) &  &  & \\
  & \left.\frac{\text{d}^2\textbf{q}}{\text{d}\textbf{x}^2}\right|_{\textbf{x}_1} &  &  &  & \textbf{C}(\textbf{x}_1) &  & \\
  &  & \ddots &  &  &  & \ddots & \\ 
  &  &  & \left.\frac{\text{d}^2\textbf{q}}{\text{d}\textbf{x}^2}\right|_{\textbf{x}_{M_H}} &  &  &  & \textbf{C}(\textbf{x}_{M_H})
\end{pmatrix}
\label{L}
\end{equation}
in which the elements in \textbf{C} are given by
\begin{equation}
\begin{split}
&C_{u,v} =J_{m,i}J_{n,j}+J_{n,i}J_{m,j}(1-\delta_{mn}) \\
&u=\frac{(i-1)(2f_x-i)}{2}+j~~(i,j\in \{1,...,f_x\},~j\ge i)\\
&v=\frac{(m-1)(2f_q-m)}{2}+n~~(m,n\in \{1,...,f_q\},~n\ge m).
\end{split}
\end{equation}
%

Finally, the training process of the transformed GPR proceeds via solving:
\begin{equation}
(\textbf{L}\textbf{K}_{qq}\textbf{L}^T+\boldsymbol{\Lambda}_{xx})\textbf{w}_\text{ext}=\textbf{y}_x
\label{tmodel}
\end{equation}
and computing $\widetilde{\textbf{w}}_\text{ext} =\textbf{L}^T\textbf{w}_\text{ext}$. 
The prediction step thus becomes,
\begin{equation}
\begin{pmatrix}
V(\textbf{x}^*)-\bar{V}\\\left.\frac{\text{d}V}{\text{d}\textbf{x}}\right|_{\textbf{x}^*}\\\left.\frac{\text{d}^2V}{\text{d}\textbf{x}^2}\right|_{\textbf{x}^*}
\end{pmatrix}
=
\begin{pmatrix}
\textbf{k}_\text{ext,q}^T(\textbf{q}(\textbf{x}^*))\\
\frac{\text{d}\textbf{k}_\text{ext,q}^T(\textbf{q}(\textbf{x}^*))}{\text{d}\textbf{x}^*}\\
\frac{\text{d}^2\textbf{k}_\text{ext,q}^T(\textbf{q}(\textbf{x}^*))}{\text{d}{\textbf{x}^*}^2}
\end{pmatrix}
\widetilde{\textbf{w}}_\text{ext}
\label{pred}
\end{equation}
in which $\textbf{k}_\text{ext,q}$ is defined similarly as $\textbf{k}_\text{ext,x}$ (Eq.~\ref{kextx}), but with the kernel derivatives taken with respect to \textbf{q} instead of \textbf{x}.
This avoids applying the complicated transformation matrix \textbf{L} in the prediction process.
Hyperparameter optimization for the coordinate transformed GPR can be done via maximizing the log marginal likelihood or via cross-validation \cite{WilliamsBook}, which is the same as that for GPR without coordinate transformation. 
%

Our transformed GPR method inherits desirable properties of the descriptor \textbf{q}.
For example, if \textbf{q} is the internal coordinates \cite{delocalizedInternals}, then one can see that the energy predictions of the GPR model (Eq.~\ref{pred}) are translationally and rotationally invariant, and that the predicted Cartesian gradients and Hessians are rotated correctly.
Obviously, Eq.~\ref{pred} avoids the potentially numerically unstable back and forth transformation of the gradients and Hessians from Cartesian coordinates to \textbf{q} coordinates, which is an advantage over the previous GPR method \cite{Laude2018}.
Furthermore, our new GPR method processes more desirable features than the previous method.
For example, since Hessian training is the most expensive and memory demanding part of GPR, it has been challenging to use GPR for instanton optimization with a high number of degrees of freedom.
The new method can train with Hessian data of selected degrees of freedom (selective Hessian training), instead of using the full Hessian, which provides a solution to the difficulty in Hessian training for large systems.
Selective Hessian training can be implemented by simply eliminating rows from \textbf{L} (Eq.~\ref{L}) that correspond to the elements discarded from the full Hessian.
Another feature is that the memory cost is now on the order of $O(f_{\textbf{y}_x}f_{\textbf{y}_q})$ instead of $O(f_{\textbf{y}_q}^2)$ ($f$ is the length, $\textbf{y}_q$ is the observables in \textbf{q} coordinates), and the computational cost is
$O(f_{\textbf{y}_x}^3)$ ($O(f_{\textbf{y}_q}^2)$ if $f_{\textbf{y}_q}>f_{\textbf{y}_x}^{3/2}$)
instead of $O(f_{\textbf{y}_q}^3)$. 
This means that the scaling of both the memory and computational costs 
with respect to length of \textbf{q} have been reduced,
making it more feasible to use longer descriptors when necessary, such as descriptors that account for permutational invariance \cite{derksen2002computational,doi:10.1063/1.4961454}.

\subsection{Descriptors for GPR learning of surface systems}
We consider systems with an adsorbed molecule (or a cluster) on a given surface.
Using Cartesian coordinates as the descriptor is a viable option, however, it is not an intrinsic descriptor for describing covalent bonds in the adsorbate.
Meanwhile, internal coordinates can describe the adsorbate well, but are not suitable as this type of systems are neither translationally nor rotationally invariant.
To describe the translation and the rotation of the adsorbate, one requires at least 6 variables, i.e. the coordinates of the centroid (\textbf{x}$_\text{c}$) and 3 rotation angles.
Intuitively, one would consider using the Euler angles, however, we found that this can be problematic.
One can imagine a simple case, a small rotation $\alpha$ about the y axis, the Euler angles (using the standard `ZXZ' convention) for this rotation is ($\pi/2$, $\alpha$, $-\pi/2$).
Under the metric of distance in the kernel (Eq.~\ref{kernel}), the Euler angles would suggest that the rotated structure is far apart from the initial structure, which is unfaithful.
This means that, at the very least, the kernel needs to be redefined to resolve this issue.
Moreover, since the adsorbate molecule is not rigid and may even dissociate in surface processes that we intend to model, Euler angles might not even be well defined and might be sensitive to the choice of the reference structure. 
Instead of trying to come up with a good descriptor for rotations, we work around this issue.
We propose an idea that combines the internal coordinates and Cartesian coordinates in order to ``gain the best of both worlds''. 
Here we use a specific definition of internal coordinates (which are the pairwise atomic distances), instead of the general definition (which includes angles and dihedrals), such that the internal coordinates have the same units as Cartesian coordinates.
First we divide the system into two parts: the adsorbates (with a total of $N_\text{a}$ atoms), and the flexible substrate atoms ($N_\text{s}$ atoms).
We also make sure that all the atoms are not wrapped by periodic boundary conditions.
Combining the Cartesian coordinates and the internals of the adsorbate gives the following descriptor,
\begin{equation}
\widetilde{\textbf{q}}=\left(\textbf{x}^\text{s};~\textbf{r};~x_0;~...~;x_{N_\text{a}};~ y_0;~...~;y_{N_\text{a}};~ z_0;~...~;z_{N_\text{a}} \right)\equiv(\textbf{x}^\text{s};~\textbf{d})
\label{descriptor0}
\end{equation}
where \textbf{x}$^\text{s}$ are the Cartesian coordinates of all the flexible substrate atoms, \textbf{d} are the coordinates representing the adsorbate and its relation with the substrate, and
\begin{equation}
\textbf{r}=(r_{ij}),~r_{ij}=||\textbf{x}_i-\textbf{x}_j||,~~i,j\in\{1,...,N_\text{a}\}~(i<j).
\end{equation}
$\widetilde{\textbf{q}}$ is obviously redundant, and it is necessary to trim out the redundancy.
We use a method similar to the method for constructing the delocalized internals from redundant internals \cite{delocalizedInternals}.
One can preform singular value decomposition on the Jacobian matrix from \textbf{x} to \textbf{d} ($\textbf{J}_{dx}$).
Since $\textbf{J}_{dx}$ is \textbf{x} dependent, previous works select a reference geometry from the training set.
Alternatively, we propose to average $\textbf{J}_{dx}$ over all the geometries in the training set, 
\begin{equation}
\textbf{USV}_h=\sum_{m=1}^M \textbf{J}_{dx}(\textbf{x}_m)/M.
\label{USV}
\end{equation}
The delocalization matrix $\textbf{B}_{qd}$ is constructed by taking the row vectors in \textbf{U}$^T$ that correspond to the non-zero singular values in \textbf{S}.
Therefore, the non-redundant descriptor is given by, 
\begin{equation}
\textbf{q}=(\textbf{x}^\text{s};~\textbf{B}_{qd}\textbf{d}).
\label{descriptor}
\end{equation}
We refer to \textbf{q} as Mixed Internals and Cartesian (MIC) descriptor. 
\textbf{q} reduces to the Cartesian coordinates when the adsorbate is a single atom.
We note that \textbf{q} is similar to an idea proposed in early years for reducing the size of the delocalization matrix for large molecules \cite{Billeter2000}. 
Our GPR PES for surface systems is built using Eqs. \ref{descriptor}, \ref{tmodel}, and \ref{pred}.
%

It is useful to discuss some of the other descriptor ideas/options that we have considered, and the reason why they were not chosen.
Firstly, we note that internal coordinates often use the $1/r_{ij}$ instead of $r_{ij}$.
It is also feasible to construct \textbf{q} (Eq.~\ref{descriptor}) using $1/r_{ij}$, however, mixing $1/r_{ij}$ with Cartesian coordinates will result in inconsistent units in the descriptor.
Another idea is to select reference points on (or near) the surface and characterize the translations and rotations using the distances of the adsorbate atoms to the reference points.
We found that the performance of this approach is sensitive to the choice of the reference points and arbitrary reference usually leads to mediocre performance, therefore we do not recommend it despite it being feasible.
%

The MIC descriptor is not limited to describing surface systems, for example, it can also be directly applied to describe reactions/processes in solids.
In this case, \textbf{x}$^\text{s}$ would represent flexible atoms in the surroundings of the core reaction region, instead of the substrate atoms.
Further, the MIC descriptor is not limited to mixing Internal and Cartesian coordinates.
One can view that the Cartesian coordinates in \textbf{d} (Eq.~\ref{descriptor0}) alternatively as coordinates describing the connection between the adsorbate and the substrate.
Thus, one can replace them with other coordinates designed to describe the connection between the core region and environment, and construct \textbf{q} (Eq.~\ref{descriptor}) from that.
Finally, the coordinates describing the environment (\textbf{x}$^\text{s}$) do not necessarily need to be Cartesian coordinates either, and one can design them according to need.
These extensions can make MIC-like descriptors useful for a wide range of systems, such as general systems that can be divided into a core region plus an environment. 

\section{Computational setup}
Electronic structure for the three test systems, namely \ce{H2O} on Cu(111), \ce{CH2O} on Ag(110), and FAD on NaCl(001), are described with density-functional theory (DFT).
Our DFT calculations were carried out using the Vienna \textit{ab initio} simulation package (VASP) \cite{vasp1,vasp2}.
The optB86b-vdW exchange-correlation functional \cite{optB86b,vdW}, which accounts for van der Waals interactions, was used. 
A plane-wave cut off of 400 eV was used (550 eV was used for the FAD system).
The Cu(111) substrate was modeled using 4-layer slab in a 3$\times$3 supercell.
The Ag(110) substrate was modeled using 4-layer slab in a 3$\times$4 supercell.
The NaCl(001) substrate was modeled using 2-layer slab in a 2$\times$2 supercell.
We used a 3$\times$3$\times$1 $k$-point mesh for all the systems. 
A vacuum of at least 12~\AA\ was placed above each slab, and a dipole corrected was applied along the z axis. 
The substrates were prepared with the top two layer relaxed (top one layer relaxed for NaCl).
The climbing image nudged elastic band (CI-NEB) method \cite{CI_NEB} was used to compute the minimum energy pathways (MEP). 
During the geometry optimization, including CI-NEB and instanton calculations, four (two) substrate atoms closest to \ce{H2O} (\ce{CH2O}) were also optimized while the other substrate atoms are kept frozen.
For FAD, the substrate was kept frozen.
The convergence criteria for geometry optimizations and CI-NEB calculations was to converge the maximum force component to below 0.02 eV$\cdot$\AA$^{-1}$.
For our instanton optimizations, the convergence criteria was to converge the \textit{total} gradient to below 0.05 eV$\cdot$\AA$^{-1}$, which is a stricter criteria than the one above.

\section{Results}
\subsection{Analysis of the descriptors}
\begin{figure}[!ht]
  \centering
  \includegraphics[width=12cm]{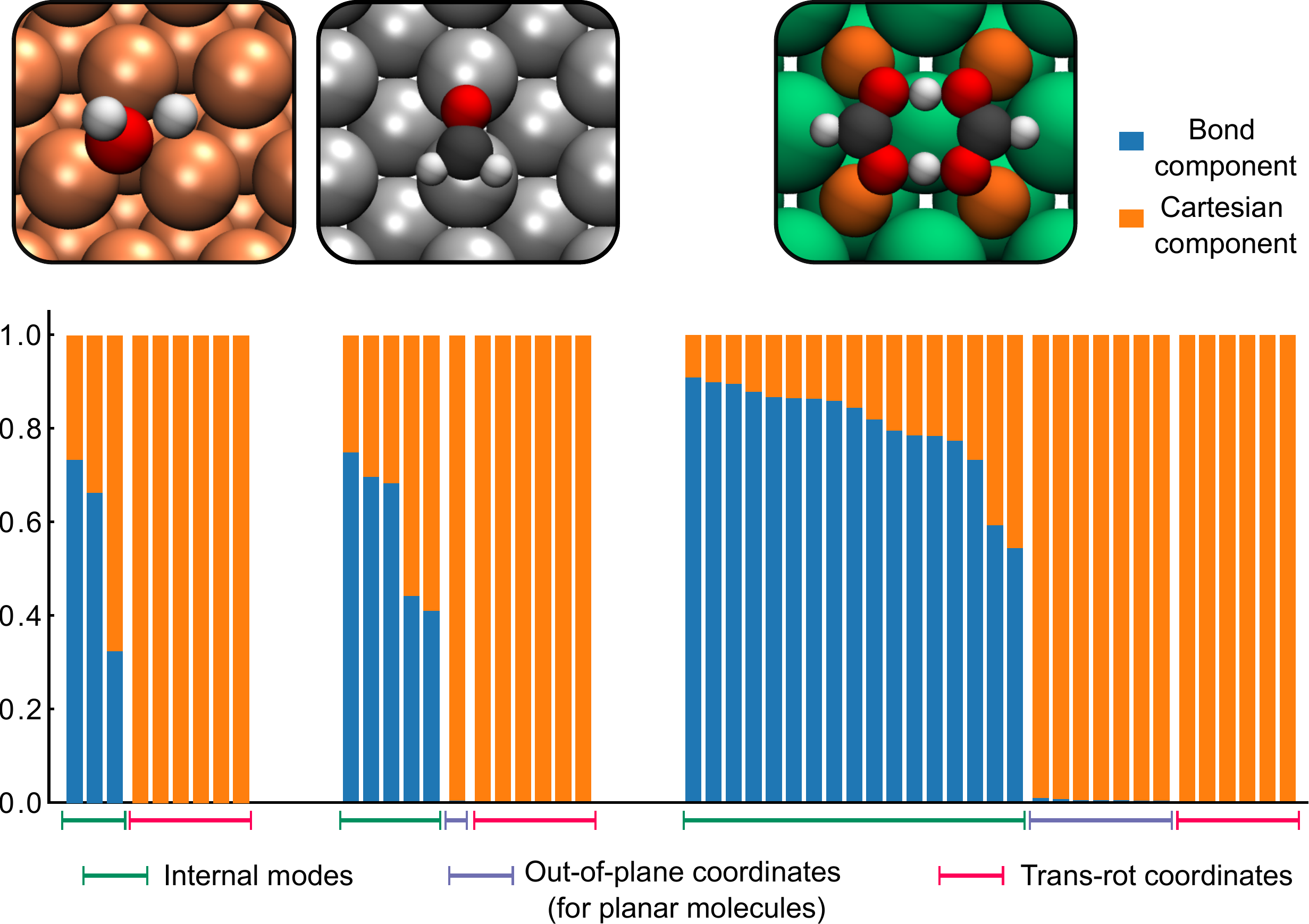}
  \caption{
  Bond and Cartesian components of MIC descriptors for the three systems.
  The transformation matrix (Eq.~\ref{USV}) is computed using geometries on the CI-NEB path.
  The bars show the unitary singular vectors (rows in \textbf{B}$_{qd}$) arranged according to their singular values in descending order.
  The bond component is the sum of the square of the first $N_a(N_a-1)/2$ terms in a singular vector, and the Cartesian component is the sum of the square of the remainder terms. 
  The top panels show the classical TS for the three systems.
  }
  \label{Jqd}
\end{figure}

First we examine our MIC descriptors for the three test systems, via a decomposition of the adsorbate-related elements in the descriptor (Fig.~\ref{Jqd}).
For \ce{H2O} on Cu(111), it has 3 non-redundant internal coordinates and 6 trans-rot coordinates.
In the MIC descriptor, the first 3 elements (which has the highest 3 singular values) have significant portion of bond components, meaning that they correspond to the internals of the molecule.
The remainder 6 elements are purely linear combinations of Cartesian coordinates, with 3 corresponding to the centroid of the molecule representing translation and the other 3 representing rotations.
For \ce{CH2O} on Ag(110), since the molecule is planar, 5 of its 6 internal coordinates are non-redundant.
Correspondingly, the first 5 elements of the MIC descriptor have significant bond components.
The 6th element consists of z coordinates, representing the out-of-plane mode of the adsorbate.
The final 6 elements represent the translation and rotation of the adsorbate.
FAD is also a planar adsorbate, so while it has $N_a(N_a-1)/2=45$ redundant internal coordinates, only $2N_a-3=17$ are non-redundant. 
Its MIC descriptor fully covers all the non-redundant internal coordinates, having 17 elements majorly consist of bond components.
MIC descriptor also has 7 elements that consist of majorly z coordinates, representing out-of-plane modes of the adsorbate.
These elements together with the 17 ``bond'' elements makes up the $3N_a-6=24$ modes representing the adsorbate, and the remainder 6 elements are the trans-rot coordinates.
The above analysis shows that our MIC descriptor not only mixed the bond and Cartesian coordinates, but also mixed them in an appropriate manner that gives a faithful description of the system.
We show later in the manuscript that it is indeed advantageous over Cartesian coordinates for GPR modeling of surface systems.

\subsection{General workflow}
\begin{figure}
  \centering
  \includegraphics[width=8.5cm]{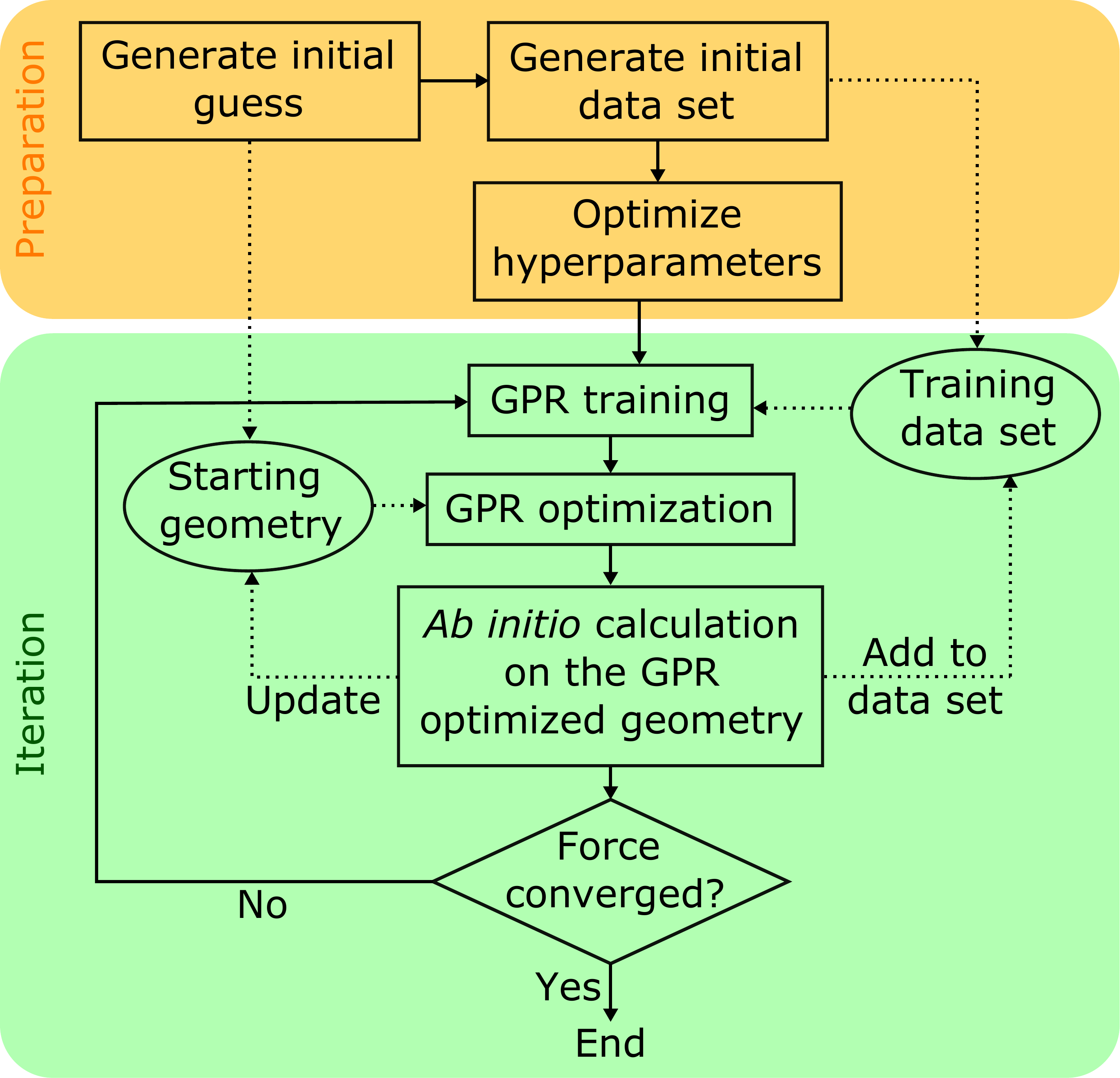}
  \caption{
  A workflow chart for GPR assisted geometry optimization.
  The dotted arrows show the flow of data (i.e. geometries and \textit{ab initio} data).
  }
  \label{workflow}
\end{figure}

In this section we describe the workflow for geometry optimization with GPR.
The whole procedure (Fig.~\ref{workflow}) is divided into two parts, ``preparation'' and ``iteration''.
Details in the preparation step differ depending on the type of calculation, i.e. instanton optimization, CI-NEB optimization, etc., and we discuss this in the next section.
Here we put hyperparameter optimization in the preparation step, meaning that we do not re-optimize them every time when we add data to the training data set in the iteration.
This is because that 
hyperparameter optimization (via log marginal likelihood) is computationally inefficient 
and we found that in the iteration process it does not change the log marginal likelihood by much.
Similar observations have been reported that the log marginal likelihood is not particularly sensitive to the hyperparameters in a reasonable range \cite{Kastner_GPR_2021}.
We also note that the workflow for GPR optimization is not unique, and one could adjust it according to the system and specific needs.
%

Using the training data set, we can build our GPR PES and perform optimizations on it.
A quasi-Newton type optimization algorithm adapted for instanton optimizations is used, which has been shown to be arguably the most stable optimization method for instantons \cite{Rommel2011locating,JeremyThesis}
Optimization proceeds until either the total force on the GPR PES reaches the convergence threshold, or an early stop condition is met.
Different early stop conditions could be defined for different applications \cite{Jonsson_GPR_2017}, in this work, an early stop is triggered if the total force increases for more than three consecutive steps.
Once early stop is triggered, we take the geometry at the step before the consecutive force increase occurred as the final geometry of the optimization step.
Since we use a quasi-Newton algorithm, early stop could be triggered due to the approximate Hessian being bad after a certain number of updates.
Therefore, when early stop is triggered for the first time, we would restart the GPR optimization once more.
After the GPR optimization, we compute the true \textit{ab initio} energy and forces on the final geometry to check whether the forces have reached our convergence criteria.
If not, we add the newly computed \textit{ab initio} data to the training set, and repeat the iteration steps.
Note that one could also perform a data refinement, which removes some of the data (e.g. data added in early iterations, or data selected based on estimates of the GPR uncertainties \cite{WilliamsBook,Jonsson_GPR_2017}) to prevent the data set becoming too large.
We find that such procedure is unnecessary for the optimizations done in this work.

\subsection{Instanton optimization}
\begin{figure}
  \centering
  \includegraphics[width=11cm]{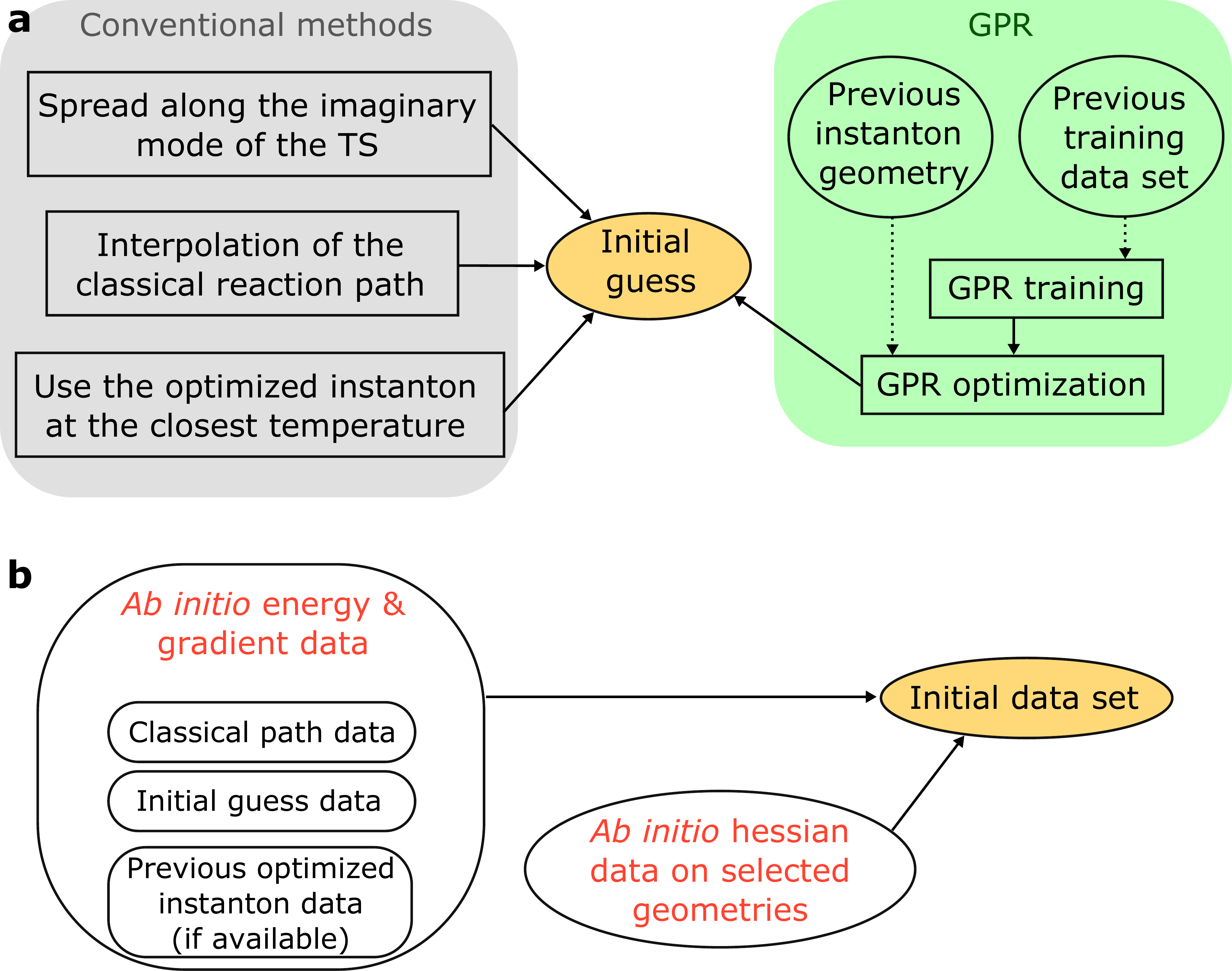}
  \caption{
  (a) Conventional methods v.s. the GPR method proposed in this work for generation initial guesses for instanton optimization.
  (b) Components of the initial data set for instanton optimization used in this work.
  }
  \label{prep}
\end{figure}

For instanton optimization, the methods proposed in this work for generating initial guess geometry and initial data set are shown in Fig.~\ref{prep}. 
Conventionally, the initial guess for instanton optimization is generated via ``spreading'' the beads along the imaginary mode of the TS \cite{Rommel2011locating}, or via interpolation of CI-NEB images (i.e. the climbing image and a few adjacent images).
If an optimized instanton configuration is available at a temperature not too far from the target temperature, then conventionally it is a good choice to start the instanton optimization from that configuration.
Therefore, if the goal is to obtain instantons at different temperatures, one would optimize them in a ``sequential cooling'' manner \cite{Rommel2011locating}, i.e. start with instanton optimization at the highest target temperature, and perform instantons optimizations in a descending order according to temperature.
We propose a GPR based approach for generating initial instanton guess when there is instanton data at another temperature available.
We train a GPR PES using the previous instanton data, and perform instanton optimizations on the GPR PES at the target temperature.
Typically, this GPR optimization could not reach the target force convergence and end via an early stop with the criteria described in the previous section.
We show later in this section that this is a good approach with the MIC descriptor, but not with the Cartesian descriptor.

After obtaining the initial instanton guess, we generate our initial data set for GPR training based on a simple protocol (as described in Fig.~\ref{prep}b).
The initial data set composes of three parts: data of the images on the CI-NEB path, data of the beads of the initial guess, and data from a previously optimized instanton at an adjacent temperature (if available).
All data points have \textit{ab initio} energy and gradient data, while \textit{ab initio} Hessian data is added for selected geometries.
In specific, the geometries with Hessian data include: the first, final, and the highest energy beads from the initial instanton guess, the first and final beads from the previous instanton (if available), and the classical TS.
Hessian data of the reactant and product geometries is also included, unless the instanton configuration is obviously far away from the reaction and product (e.g. for \ce{H2O} dissociation on Cu).
We note that there are other more complicated protocols for generating the initial data set and for selecting Hessian data, which could be explored in the future.
\begin{table}
  \caption{
  Settings and initial data set components, i.e. the number of energy and gradient data points (n$_\text{ener,grad}$) and the number of Hessian data (n$_\text{hess}$), for each GPR assisted instanton optimization.
  The crossover temperature to quantum tunneling, estimated by $T_c\sim\frac{\hbar |\omega^{\ddagger}|}{2\pi k_\text{B}}$, where $\omega^{\ddagger}$ is the imaginary frequency at the TS, is also given for each system.
  }
  \label{initset_size}
  \begin{tabular}{c|cc|cc}
    \hline
    \hline
    system  & $T$ (K) & $N$ beads & n$_\text{ener,grad}$ & n$_\text{hess}$  \\
    \hline
    \ce{H2O}-Cu(111) & 200 & 14 & 16 & 3 \\
     ($T_c\sim281$ K)    & 130 & 30 & 31 & 6 \\
                     &  80 & 50 & 49 & 6 \\
    \hline
    \ce{CH2O}-Ag(110) & 18 & 14 & 14 & 5 \\
     ($T_c\sim22$ K)    & 12 & 30 & 29 & 8 \\
                      &  8 & 50 & 47 & 8 \\
    \hline
    FAD-NaCl(001) & 150 & 14 & 14 & 5 \\
     ($T_c\sim223$ K)   & 100 & 30 & 29 & 8 \\
    \hline
    \hline
  \end{tabular}
\end{table}

We give the detail settings for our GPR assisted instanton optimizations in Table \ref{initset_size}.
The aim of this work is the to demonstrate the performance of our method for different types of system from the shallow tunneling to deep tunneling regimes.
Therefore, three temperatures are chosen for each system (except for FAD) such that at the highest temperature, thermally activated tunneling near the barrier top occurs; at the middle temperature, tunneling through the middle of the barrier occurs; and at the lowest temperature, deep tunneling occurs.
Since computing fully converged instanton rates for these test systems is not the aim of this work, hence at each temperature, we use the minimal number of beads that can reasonably represent the tunneling pathway. 
In the next section, we show how GPR can be used to extrapolate instantons from a small number of beads to a large number of beads.
The sizes of the initial data sets generated with the protocol in Fig.~\ref{prep}b are given in Table~\ref{initset_size}.
One can see that the sizes are quite modest and do not vary much for different systems.
%

\begin{figure}[!ht]
  \centering
  \includegraphics[width=12cm]{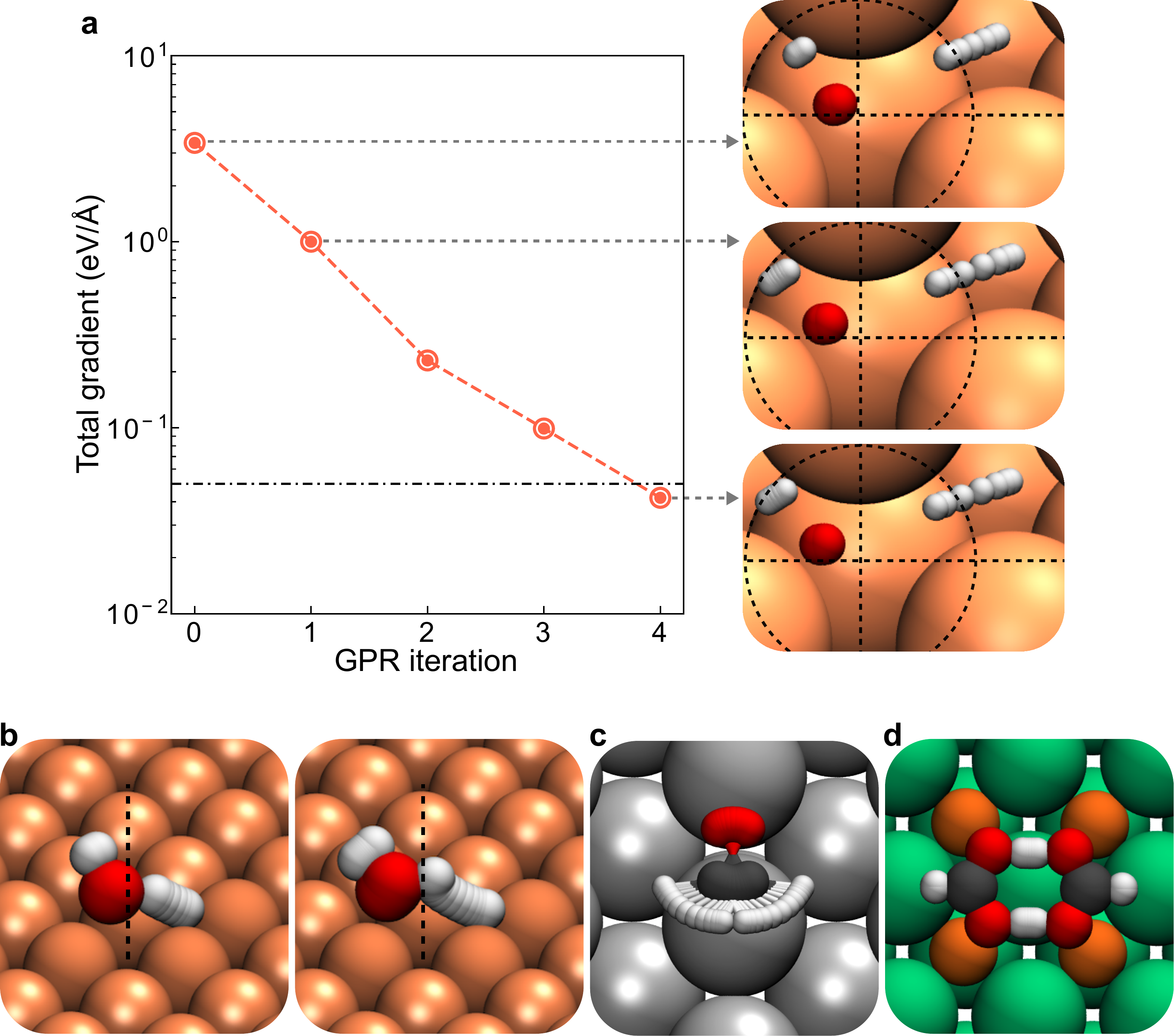}
  \caption{
  (a) Instanton geometries for dissociation on Cu(111) at 200 K at the 0th (initial guess), 1st, and final iteration during the GPR assisted optimization.
  Norm of the \textit{ab initio} total gradient during the optimization process is shown, and the dash-dotted line marks the convergence criteria.
  Optimized instanton geometries for \ce{H2O} dissociation on Cu(111) (b) at 200 K (left) and 80 K (right), \ce{CH2O} rotation on Ag(111) at 12 K (c), and DPT in FAD on NaCl(001) at 100 K (c).
  The flexible surface atoms in (b) and (c) are rendered in a different texture than the other substrate atoms. 
  The dotted-lines are guides for the eye.
  }
  \label{inst_geo}
\end{figure}

Using the procedures described above, we preformed \textit{ab initio} instanton optimizations for the test systems at the different temperatures in a ``sequential cooling'' manner.
Note that the temperature gap between adjacent instanton optimizations are large, corresponding to a $\beta$ increase of $\sim$ 50\%. 
Encouragingly, all the GPR assisted instanton optimizations successfully converged with ease, demonstrating that our GPR method is able to model all the three types of process: dissociation on surface, rotation on surface involving heavy atom tunneling, and proton transfer between adsorbates.
Examples of the optimized instantons are shown in Fig.~\ref{inst_geo}.
Noticeably, even when the instanton displays significant corner-cutting effects (Fig.~\ref{inst_geo}b indicated by the change in the O position in the two instanton), our GPR method still performs well.
To gain a straight-forward view of the GPR assisted optimization process, we show in Fig.~\ref{inst_geo}a the geometry and total \textit{ab initio} gradient after each GPR iteration (Fig.~\ref{workflow}).
The total gradient decreases exponentially after each GPR iteration, which is a sign of efficient and stable optimization.
A feature of GPR optimization is that unlike conventional methods, it allows relatively major changes in the geometry after one iteration, such as in the first GPR iteration shown in Fig.~\ref{inst_geo}a, without worrying much about stability issues.
%

As the contrast, we performed \textit{ab initio} instanton optimizations for these systems using conventional methods.
There are a few conventional optimization algorithms that have been adapted for \textit{ab initio} instanton optimization, belonging to two categories:
mode following methods (i.e. dimer methods) \cite{dimer,superdimer}, and
Hessian based methods (i.e. quasi-Newton methods) \cite{Nichols1990mep,Bofill1994update,JeremyThesis}.
The performance of these instanton optimization algorithms have been discussed in detail in previous works \cite{Rommel2011locating,JeremyThesis}.
In short, Hessian based methods are more stable and converge faster than mode following methods, but require calculation of the Hessians for the starting configuration, as they tend to fail for instanton optimization without a good estimate of the initial Hessian.
This means that for systems with a limited number of flexible atoms (i.e. $\le$10), Hessian based methods are overall more efficient, while for systems with many flexible atoms, mode following methods can be advantageous. 
Since the test systems have relatively small number of flexible atoms, we used the quasi-Newton method described in Ref. \citenum{JeremyThesis} as the conventional instanton optimization method.
\begin{table}
  \caption{
  Comparison of the convergence speed (i.e. the number of iterations n$_\text{iter}$) and computational cost between \textit{ab initio} instanton optimization with GPR assistance and that with conventional quasi-Newton algorithm.
  n$_\text{hess}$ is the number of \textit{ab initio} Hessian evaluations preformed.
  The superscript `*' means that the \textit{ab initio} optimization required restarting from a selected intermediate geometry in the optimization process, and re-computing the Hessian on this geometry.
  In this case, the number of iterations actually performed is larger than n$_\text{iter}$.
  }
  \label{cost_cmp}
  \begin{tabular}{c|cc|cc|cc}
    \hline
    \hline
    system  & $T$ (K) & $N$ beads & \multicolumn{2}{c|}{GPR assisted} & \multicolumn{2}{c}{Conventional} \\
    & & & ~~~n$_\text{iter}$~~ & ~~n$_\text{hess}$~~~ & ~~~n$_\text{iter}$~~ & ~~n$_\text{hess}$~~~ \\
    \hline
    \ce{H2O}-Cu(111) & 200 & 14 & 4 & 2 & 18 & 7 \\
                     & 130 & 30 & 4 & 5 & 15$^*$ & 15 \\
                     &  80 & 50 & 6 & 5 & 18$^*$ & 40 \\
    \hline
    \ce{CH2O}-Ag(110) & 18 & 14 & 2 & 2 & 19$^*$ & 14 \\
                      & 12 & 30 & 1 & 5 & 29$^*$ & 22 \\
                      &  8 & 50 & 1 & 5 & 27 & 15 \\
    \hline
    FAD-NaCl(001) & 150 & 14 & 2 & 2 & 7 & 7 \\
                   & 100 & 30 & 1 & 5 & \multicolumn{2}{c}{Not converged} \\
    \hline
    \hline
  \end{tabular}
\end{table}
%

A full comparison of the convergence of \textit{ab initio} instanton optimization with GPR and with the conventional method for all the 8 instanton optimizations is presented in Table~\ref{cost_cmp}.
The results show that in all the cases tested, our GPR method clearly out-performs the conventional method for instanton optimization in surface reactions. 
With our GPR method, all these instantons appear to be very easy to optimize, 
while for the conventional method, the opposite is true, as the majority of the optimizations takes many iterations.
In several cases, conventional optimization method becomes unstable after some iterations and require a restart from a selected intermediate configuration (as well as re-computing the initial Hessian) in order to converge the optimization.
This means that the actual number of \textit{ab initio} energy and force evaluations on the instanton in these cases are even large than the n$_\text{iter}$ given in the table.
Overall, one can see that our GPR method reduces the cost of \textit{ab initio} instanton optimizations on surfaces by around a factor of 5 or even more.

\begin{figure}
  \centering
  \includegraphics[width=10cm]{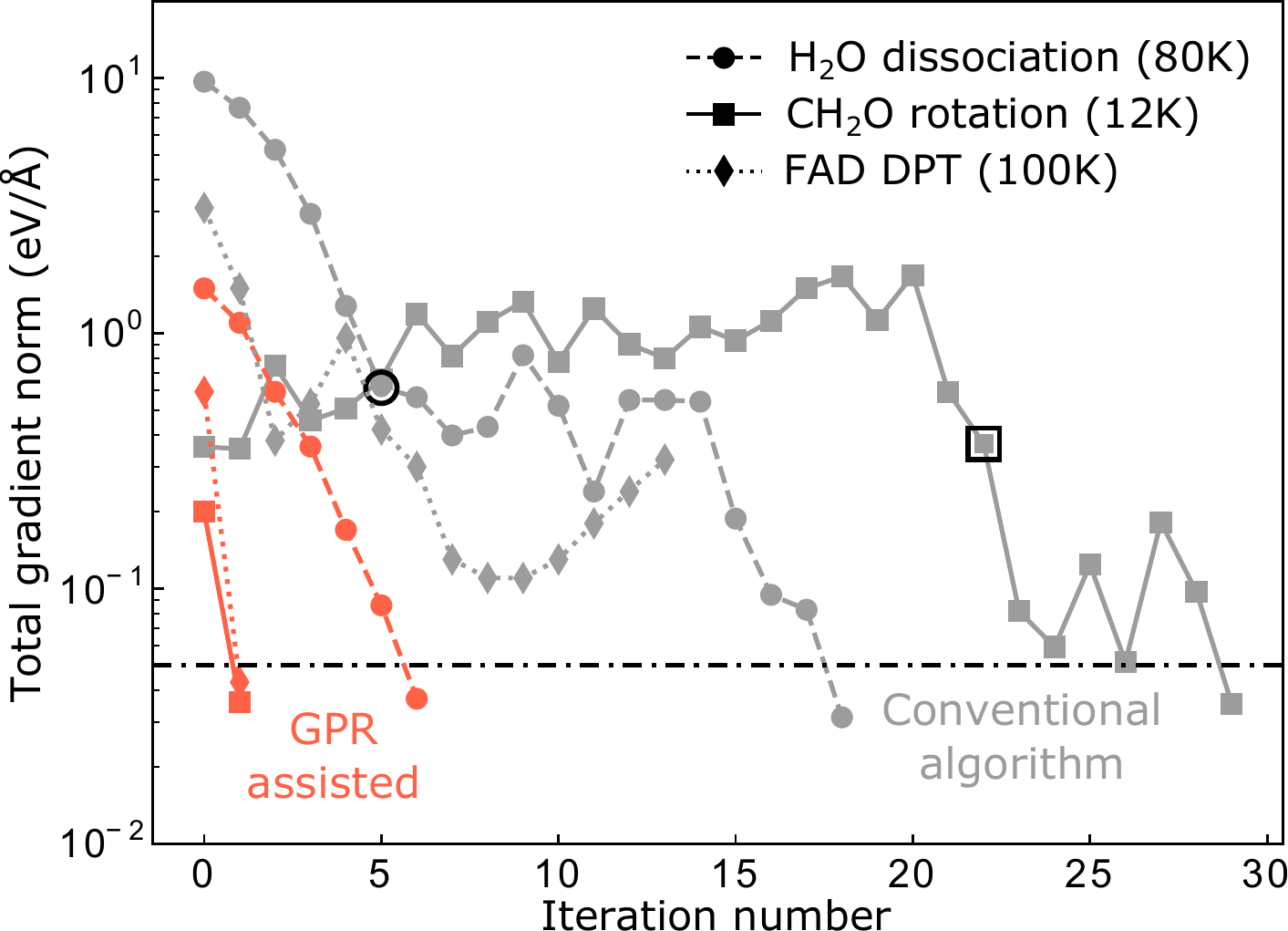}
  \caption{
  Norm of the total gradient at each iteration for selected \textit{ab initio} instanton optimization with GPR assistance (red) and with conventional algorithm (grey). 
  The circled points mark the restart points of the \textit{ab initio} instanton optimization.
  }
  \label{conv}
\end{figure}

To understand why GPR drastically outperforms the conventional method, 
we compare the change of the total gradient during the entire optimization process for the two methods (Fig.~\ref{conv}).
In the conventional instanton optimization for \ce{H2O} dissociation and FAD DPT, during the first few iterations the total gradient on the instanton decreases quite rapidly.
However, afterwards the approximate Hessian becomes poor due to accumulation of errors from the updates. 
As a result, the optimization fails to further minimize the total gradient and a restart with a re-evaluation of the instanton Hessian (which can be computationally demanding) is needed.
Even after restarting, the optimization could still fail, such as for FAD (which we suspect due to instabilities caused by two soft translational modes).
At the core of this issue is that conventional methods barely learn anything from the \textit{ab initio} data computed in previous iterations, wasting a lot of useful information and resulting in taking misguided steps.
GPR exploits the previous data in-depth, learning useful information that provides guidance for each optimization step taken, which greatly accelerates the convergence.
%

\begin{figure}
  \centering
  \includegraphics[width=12.5cm]{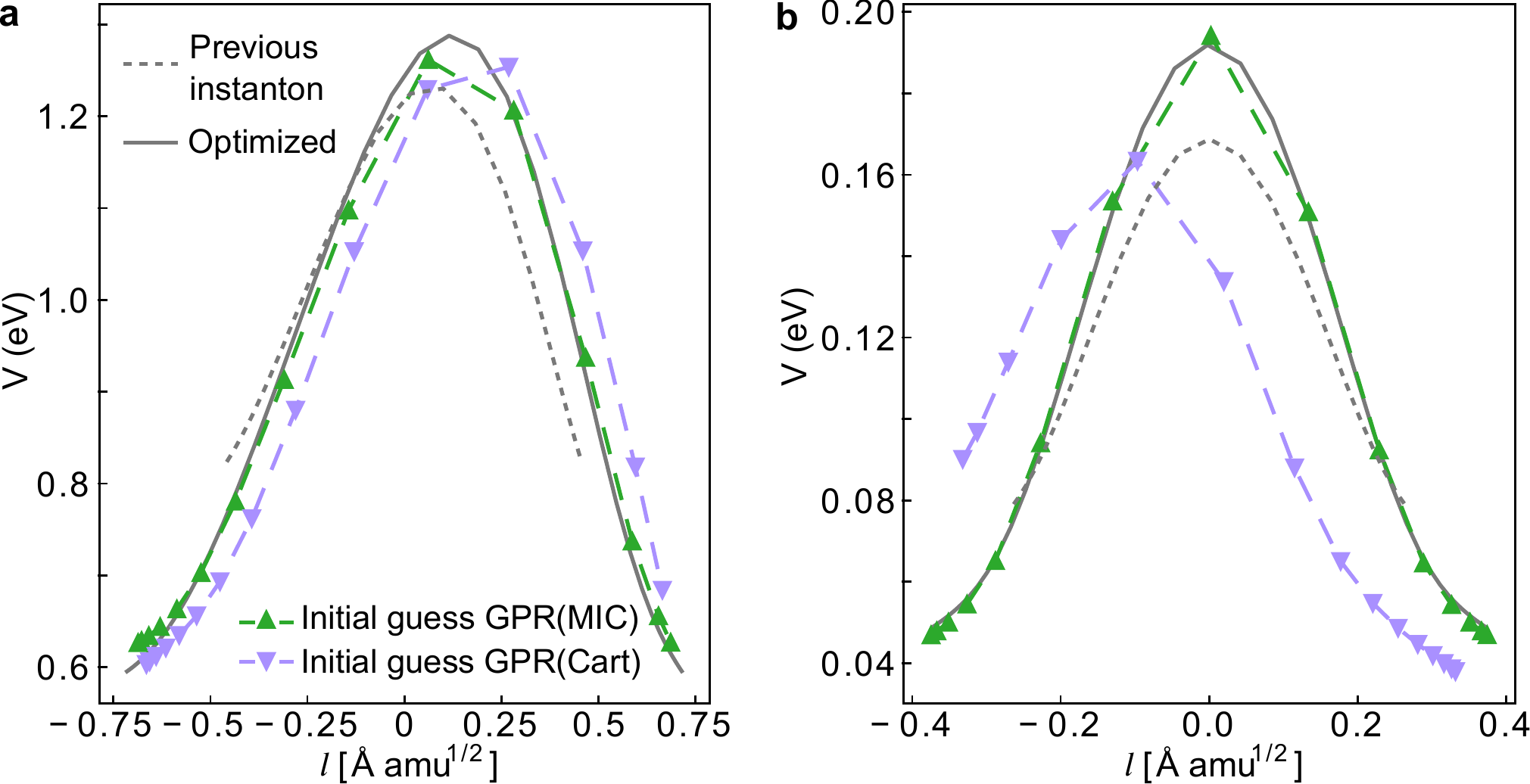}
  \caption{
  Comparison of the initial instanton guess predicted by MIC GPR PES and that predicted by Cartesian GPR PES for (a) \ce{H2O} dissociation on Cu(111) at 130 K and (b) DPT in FAD on NaCl(001) at 100 K.
  The ring-polymer beads are plotted as a function of their potential energy and the path length $l$.
  For reference, the previous instanton (grey dotted line) from which the GPR PESs are trained and the final optimized instanton (grey solid line) are shown.
  }
  \label{cart_vs_MIC}
\end{figure}

The in-depth learning power of our GPR method gives it another key advantage, the ability to generate a good initial guess that is very close to the optimized instanton, using only instanton data accumulated in the previous instanton optimization.
This is indicated in Fig.~\ref{conv}, that the GPR initial guesses have significantly smaller \textit{ab initio} gradient compared to the conventionally used initial guesses, which is another important reason why GPR assisted optimization converged very fast.
Here we demonstrate that the in-depth learning is achieved by the use of MIC descriptor, while using Cartesian descriptor will produce much worse predictions.
A plot of the GPR (with MIC descriptor) initial guess and the optimized instanton (green vs grey lines in Fig.~\ref{cart_vs_MIC}) shows that they are almost identical.
This is particularly exciting 
given that machine learning methods are generally not good at ``extrapolation'', and the fact that corner-cutting occurs in two of the systems tested (see the dotted vs solid grey lines in Fig.~\ref{cart_vs_MIC}). 
We attribute this to two factors, first is Hessian learning, which have been shown in previous works to be important for GPR optimization \cite{Jonsson_GPR_2017,Laude2018}, and more importantly, that our MIC descriptor provides a faithful description of surface systems.

In contrast, GPR with the Cartesian descriptor does not work well, predicting in a much worse initial guess for instanton optimization (see Fig.~\ref{cart_vs_MIC}).
For \ce{H2O}, by examining the geometries, we find that in the Cartesian GPR initial guess, the O atom does not translate much from the previous instanton, indicating that Cartesian descriptor could not predict corner-cutting effects well.
This is expected as the Cartesian descriptor does not see the adsorbate as a molecule, hence does not recognize the similarity of it before and after translation.
The MIC descriptor does not suffer from this problem, hence MIC GPR can learn the data in depth for surface systems, whereas Cartesian GPR can only learn it shallowly.
When the adsorbate is larger, for example for FAD, Cartesian GPR fails more drastically and could not even predict a remotely reasonable initial instanton guess (Fig.~\ref{cart_vs_MIC}).
Even for systems in which no corner-cutting is observed, i.e. \ce{CH2O} rotation, we find that MIC still out-preforms Cartesian descriptor, predicting an initial instanton guess that is slightly closer to the optimized instanton than that predicted with Cartesian GPR.
Therefore, we conclude that MIC offers an intrinsic and faithful description of surface systems, which we recommend over the Cartesian descriptor.

\subsection{Selective hessian training}
Finally, as a proof of concept, we demonstrate the performance of selective Hessian training on two systems, \ce{H2O} dissociation and DPT in FAD.
A straight-forward way to select Hessian elements for surface systems is to divide them into three parts, i.e. the adsorbate part, the substrate part, and the adsorbate-substrate part, and select the desired parts.
For the \ce{H2O} dissociation instanton optimization at 130K, the initial data set has 6 Hessians, which (as describe previously) correspond to the classical TS, the end beads of the instanton at 200K, and the first, middle (highest energy), and final beads of the initial instanton guess.
We select only the adsorbate part for the first three Hessians, and for the latter three Hessians we discard the substrate part of the Hessian on the middle bead.
This reduces the size of the Hessian data by 46\%.
Using this reduced training set, we re-performed the GPR assisted instanton optimization, and compared the performance to GPR optimization without selective Hessian training.
The results are encouraging (see Table \ref{selhess}), the optimization takes only 4 iterations to converge, which is the same as using GPR with full Hessians.
In addition, we tested that it also works well for generating initial instanton guess, predicting almost identical results as GPR with full Hessians.
\begin{table}
  \caption{
  Convergence speed (i.e. the number of iterations n$_\text{iter}$) and computational cost of the selective Hessian training approach.
  n$_\text{hess}$ is the same as in Table~\ref{cost_cmp}.
  ``Reduction'' shows the percentage reduction in the size of the Hessian data in the training set with this approach.
  }
  \label{selhess}
  \begin{tabular}{c|cc|ccc}
    \hline
    \hline
    system  & $T$ (K) & $N$ beads & ~~~n$_\text{iter}$~~ & ~~n$_\text{hess}$~~~ & reduction \\ 
    \hline
    \ce{H2O}-Cu(111) & 130 & 30 & 4 & 5 & 46\% \\
    \hline
    FAD-NaCl(001) & 150 & 14 & 2 & 2 & 26\% \\
    (flexible substrate)   & 100 & 30 & 1 & 5 & 41\% \\
    \hline
    \hline
  \end{tabular}
\end{table}
For the FAD on NaCl(001) system, we re-preformed the instanton optimization with 7 flexible surface atoms (closest to the adsorbate).
This is very expensive to preform with conventional methods.
With the GPR + selective Hessian approach, we are able to converge the instanton with minimal computational effort, i.e. 1 or 2 iterations and only a few Hessian calculations (Table \ref{selhess}).
These systems tested pose no challenge for selective Hessian training, showing that this approach (when done properly) does not compromise performance at all, meanwhile reduces the size of the training set considerable, revealing the great application potential of our GPR optimization approach in larger systems.

\subsection{Converging instanton rates}
In the previous section, we demonstrate instanton optimization with GPR, next we discuss how GPR can be used to obtain converged instanton rates.
The instanton rate is given by: 
\begin{equation}
    k_{\text{inst}}(\beta;N) = A_{\text{inst}}(\beta;N) \, \mathrm{e}^{-S(\beta;N)/\mathrm{\hbar}},
    \label{k_inst}
\end{equation}
in which $S$ is the Euclidean action of the instanton (represented by a $N$-bead ring-polymer) with imaginary time $\beta\mathrm{\hbar}$ ($\beta=1/k_{\text{B}}T$), and $A_{\text{inst}}$ is a measure of fluctuations around the instanton path \cite{Inst_persp,InstReview}.
Instanton rates need to be converged with respect to the number of beads, and 
at low temperatures, a large number of beads is often required.
There are two approaches to converge instanton rate calculations with GPR: 
the ``rigorious'' approach is to use GPR optimization to obtain the instanton with a large number of beads and compute the instanton rate on it; 
the ``approximate'' approach is to train GPR using the data of the instanton with a small number of beads and compute instanton rates with a large number of beads on the GPR PES.
We demonstrate using both approaches to compute the instanton rate for \ce{H2O} dissociation on Cu(111) at 80 K.
For the first approach, the most computationally efficient way to obtain an optimized instanton with a large number of beads is to first optimize the instanton with a small number of beads, and then starting from this optimize the instanton with a large number of beads.
Using this approach, we optimized the instanton with 130 beads, and computed \textit{ab initio} instanton rate on this configuration rigorously.
This rate serves as the benchmark for the ``approximate'' approach.
Despite with the assistance of GPR optimization, computing fully converged instanton rates rigorously can still be computationally expensive, especially at low temperatures, due to the large number of beads required.
Here we explore the accuracy of instanton rates computed on the GPR PES, which is computationally inexpensive, to test whether they can be used as a good approximation to \textit{ab initio} instanton rates computed rigorously.
Note that previous study showed that GPR rates for reactions like \ce{CH4}+H do agree well with rigorous instanton rates \cite{Laude2018}.
The data set for GPR training contains the energy and gradient data for all the beads in the optimized instanton (25 geometries as the instanton folds on itself).
Hessian data on selected geometries (ensuring that these geometries are roughly evenly spaced along the instanton path) are also added.
%

\begin{table}
\caption{
Comparison of the instanton rate computed on the GPR PES and the \textit{ab initio} instanton rate for \ce{H2O} dissociation on Cu(111) at 80 K. 
The \textit{ab initio} instantons were optimized with GPR assistance.
The tunneling factor is defined as the ratio of the $N$ bead instanton rate and the Eyring TST rate in the $N$ bead limit.
}
\label{rates_cmp}
\centering
\begin{tabular}{l|ccc|ccc|c}
\hline
\hline
PES & \multicolumn{1}{c}{$N$} & \multicolumn{1}{c}{n$_\text{hess}$} & \multicolumn{1}{c|}{n$_\text{ener,grad}$} & \multicolumn{1}{c}{$S/\hbar$} & \multicolumn{1}{c}{$k_\text{inst}$ (N)} & \multicolumn{1}{c|}{tunneling factor} & error (\%)  \\
\hline
DFT & 50    & -      & -         & 116.032      & 5.11E-31     & 5.47E+22 &-\\
GPR  & 50     & 7       & 25           & 116.030      & 3.03E-30     & 3.25E+23  & $>$100 \\
GPR  & 50     & 10       & 25          & 116.030      & 4.33E-31     & 4.64E+22  & 15.2 \\
DFT & 130  & -      & -          & 116.144      & 1.85E-30     & 3.37E+22  &- \\ 
GPR  & 130   & 10      & 25            & 116.156      & 1.44E-30     & 2.62E+22   & 22.2 \\
GPR  & 130   & 10      & 90            & 116.157      & 1.86E-30     & 3.40E+22  & 0.7 \\
\hline
\hline
\end{tabular}
\end{table}

We compared ``approximate'' instanton rates (computed on GPR PES) with \textit{ab initio} results (Table~\ref{rates_cmp}).
A first thing to note that for this reaction, at 80 K, using 50 beads is clearly inadequate, resulting in a factor of $\sim$4 error in the rate and $\sim$70\% error in the tunneling factor compared to the result with 130 beads.
Secondly, the accuracy of the GPR rate is sensitive to the number of Hessian data in the training set, as one can see, using 7 Hessians result in large errors of over 100\%, where as using 10 Hessians the error decreases to $\sim$15\%.
The instanton rate computed on the GPR PES with 130 beads has an error of $\sim$20\% compared to the benchmark, which is acceptable as this is comparable to the error of instanton theory itself.
In principle, adding more \textit{ab initio} data to the training set can further reduce the error (to only $\sim$1\% in this example), but might not be necessary.
We note that the Cartesian descriptor also performs very well for this purpose, 
since the GPR PES only needs to learn about information confined to an already identified path.
The good performance of the ``approximate'' approach has significant implications in that the computational cost of GPR assisted instanton rate calculation is now comparable to a classical TST rate calculation.
If we assume that the Hessian calculation is the bottleneck, the a TST calculation requires 2 to 3 Hessians.
With the help of GPR, an instanton calculation takes 3-8 Hessians for the optimization, and $\sim$10 Hessians (depending on the temperature, the higher the fewer) for obtaining the rate.
In contrast, instanton calculation carried out conventionally would be $\sim$2 orders of magnitude more expensive than TST, e.g. for the system demonstrated it would need over 100 Hessian calculations.
This means that performing a GPR instanton calculation is less than a order of magnitude more expensive than TST, while being orders of magnitude closer to the correct result for reactions where quantum tunneling play an important role.

\section{Conclusions}
We have proposed a robust method for GPR assisted geometry optimization with general descriptors and an improved descriptor (MIC) over Cartesians for surface systems.
The new GPR method has several advantages over the previous method, including that it no longer preforms transformations of physical observables, which avoids associated numerical issues.
MIC descriptor provides a more intrinsic and faithful description of surface systems, thus improving the performance of GPR.
\textit{Ab initio} instanton optimizations for surface reactions can be made efficient, even for cases where significant corner-cutting occurs, using MIC-GPR to fit the PES locally around the tunneling path. 
This is demonstrated using three example systems representing different type of surface reactions and processes.
GPR assisted instanton optimization can obtain converged instantons with just a few iterations, truly a significant speed up from conventional optimization methods.
This method brings down the cost for performing an instanton calculation to within an order of magnitude of the cost of a classical TST calculation, meaning that if TST is affordable, there's no reason not to perform an instanton calculation if there might be tunneling effects.
We attribute the good performance to the MIC-GPR method achieving in-depth learning of the data generated in the optimization process. 
%

We believe that the method proposed in this work has extensive application potential beyond what we have explored here.
The MIC descriptor is obviously applicable beyond surface systems, e.g. it can be used to describe processes in molecular crystals, offering a better alternative to the Cartesian descriptor. 
Our new GPR framework for general coordinates can also alleviate some of the issues/challenges that GPR optimization face.
Most noticeably, selective Hessian training under the new framework can allow GPR optimization to be applied to large systems, and we have demonstrated that it reduces the cost while preserving good performance.
We expect that GPR based optimization schemes would replace conventional methods for the more difficult optimization calculations in the future.
We are optimistic that the developments in this work bring us towards easy computation of reaction and tunneling pathways.
%

\begin{acknowledgement}
The authors thank Prof. M. Ceriotti, Mr. G. Laude, and Mr. T.-H. Jiang for the helpful discussions on GPR, and Prof. X.-Z. Li for the helpful advice on the project.
We would like to thank Mr. D. Prakash for testing some early ideas that led to the design of the method presented in this paper.
W.F. and J.O.R. acknowledge financial support from the Swiss National Science Foundation through Project 207772.
%
%
The simulations were performed on the supercomputer in Peking University and the ETH Euler cluster.
\end{acknowledgement}




\providecommand{\latin}[1]{#1}
\makeatletter
\providecommand{\doi}
  {\begingroup\let\do\@makeother\dospecials
  \catcode`\{=1 \catcode`\}=2 \doi@aux}
\providecommand{\doi@aux}[1]{\endgroup\texttt{#1}}
\makeatother
\providecommand*\mcitethebibliography{\thebibliography}
\csname @ifundefined\endcsname{endmcitethebibliography}
  {\let\endmcitethebibliography\endthebibliography}{}

\end{document}